\documentstyle[epsfig]{mn}
\def\gsimeq{{_>\atop^{\sim}}}
 
\begin{document}

\title[]{The radio luminosity density of star-burst galaxies and co-moving
star formation rate}
\author[Seb Oliver, Carlotta Gruppioni, Stephen Serjeant]{Seb Oliver, 
Carlotta Gruppioni, Stephen Serjeant
\\
Astrophysics Group,\\ Blackett Laboratory,
\\Imperial College of Science Technology \& Medicine,\\Prince Consort
Rd.,\\London.\\SW7 2BZ\\
email: s.oliver@ic.ac.uk, cgrup@ic.ac.uk, s.serjeant@ic.ac.uk}
\date{Accepted ;
      Received ;
      in original form }

\pagerange{\pageref{firstpage}--\pageref{lastpage}}
\pubyear{1998}
\volume{999}

\label{firstpage}

\maketitle


\begin{abstract}

We present a new determination of the co-moving star formation density
at redshifts \mbox{$z\stackrel{<}{_\sim}0.35$} from the $1.4$ GHz
luminosity function of sub-mJy star-burst galaxies. Our sample, taken
from Benn et al. (1993), is insensitive to dust obscuration.  The
shape of the Luminosity function of this sample is indistinguishable
from a number of reasonable a prior models of the luminosity function.
Using these shapes we calculate the modest corrections (typically
$\stackrel{<}{_\sim}20$ per cent) to the observed $1.4$ GHz luminosity
density. We find that the cosmic variance in our estimate of this
luminosity density is large.  We find a luminosity density in broad
agreement with that from the RSA sample by Condon {\em et al.} (1987).
We infer a co-moving star formation rate surprisingly similar to coeval
estimates from the Canada-France Redshift Survey, in both ultraviolet
and H$\alpha$, although the later may also be affected by Cosmic
variance.  We conclude that the intermediate $0.05<z<0.3$
star-formation rate is not yet well determined partially due to
uncertain extinction corrections, but also partially due to cosmic
variance.  We suggest that deep moderate area radio surveys will improve this
situation considerably.

\end{abstract}


\maketitle
\section{Introduction}
In recent years a major focus of extra-galactic astrophysics has
been in estimating the evolving, volume-averaged star formation rate. 
This work has been particularly stimulated by
major advances in our ability to identify star forming galaxies
at high redshifts with optical colour techniques (see e.g. Pettini et
al 1998 and references therein).  Early indications from
comparisons of such populations in the Hubble Deep Field with
similar populations at lower redshift suggested that there may
have been a peak in star formation activity at a redshift of
$z\sim 2$ (Madau et al 1996).

The theoretical interest in such measurements
is high, partly because the integrated star formation is the major
uncertainty in evolutionary models used to explain (amongst other
things) the local chemical enrichment in our own galaxy,  but also
because models of galaxy formation based on e.g. hierarchically
clustering scenarios 
can now make specific predictions for the integrated star formation rate at
different epochs (e.g. Pei \& Fall, Baugh et al 1997).

A variety of techniques have been employed to estimate the star
formation rate at a number of epochs. The common feature of such
techniques is to identify a tracer of star formation activity and
integrate this over volume, correcting for contributions missed
through sensitivity limits, obscuration or other incompleteness
effects.  Tracers that have been employed, include: the
B-band-luminosity (Lilly et al.) U-band luminosity (Madau et al 1996),
the $H\alpha$ luminosity (Kennicutt 1983), the mid-IR luminosity
(Rowan-Robinson et al 1997). The optical tracers are potentially
significantly affected by dust-obscuration and the corrections for
this are uncertain and may be redshift dependent.  In contrast Mid and
Far-IR luminosity traces the star formation rate in obscured regions
only. The obscured fraction itself is hotly debated. For example,
de-reddening of optical-UV spectra does not guarantee an unbiased star
formation rate, since the most heavily obscured and potentially
dominant star forming regions may be wholly undetected in the UV.

In this paper we present the first estimate of the co-moving star
formation density using as a tracer the radio luminosity of star
forming galaxies.  The radio luminosity traces the supernovae
associated with star formation regions, and is not significantly
affected by dust obscuration.

The paper is organised as follows. In the first section we describe
our choice of radio sample.  In the second we describe our estimation
of the radio-luminosity density including corrections for
incompleteness.  In the third section we describe the conversion from
luminosity density into star formation rate.  In the final section we
discuss our estimation of the star formation rate in the context of
other determinations and other implications.

\section{The 1.4GHz  Sample}

Deep 1.4 GHz radio source counts obtained in the last decade (Condon
\& Mitchell 1984; Windhorst 1984; Windhorst et~al. 1985) show a
steepening below a few mJy.  This excess of faint sources relative to
the Euclidean value can neither be attributed to the giant ellipticals
and quasars responsible of most of the counts observed at fluxes $> 5$
mJy, nor to the non-evolving population of normal spiral and Seyfert
galaxies.  Optical identification works published so far have shown
that the sub--mJy sources are mainly identified with faint blue
galaxies (Kron, Koo \& Windhorst 1985; Thuan \& Condon 1987), often
showing peculiar optical morphologies indicative of interaction and
merging phenomena and spectra similar to those of the star--forming
galaxies detected by IRAS (Benn et~al. 1993, B93). Although all these
works are based on very small percentages of identification (due to
the fact that the majority of the faint radio sources have very faint
optical counterparts), they all agree that most of the sub--mJy radio
sources with identifications brighter than $B \simeq 22-22.5$ are
star--forming galaxies.  In contrast, recent extensions to fainter
optical magnitudes show strong hints of an increase in the fraction of
early--type galaxies among the identifications of sub--mJy sources
($S_{1.4~GHz} \geq 0.2$ mJy; Gruppioni et al. 1998, in
preparation). Thus, down to optical magnitudes of $B \sim 22.5$ most
(or eventually all) of the star-burst counterparts of sub--mJy sources
brighter than $S_{1.4~GHz} \sim 0.1-0.2$ mJy should be detected.

\begin{figure}
\epsfig{file=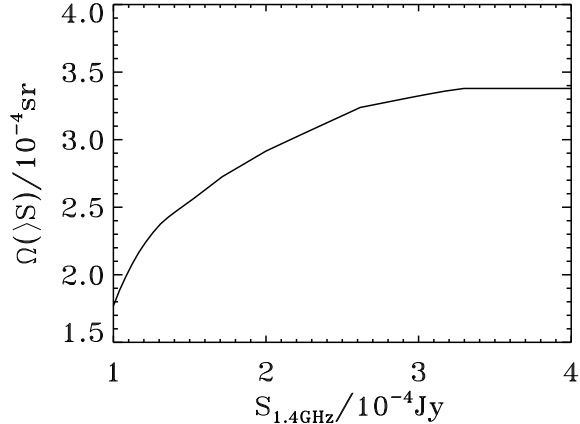, angle=90, width=8cm}
\caption{The areal coverage $\Omega(S)$ of the B93 sample as
a function of 1.4 GHz flux limit}\label{fig:omega}
\end{figure}

The B93 identification sample consists of 112 candidate optical
counterparts of $\sim500$ sub--mJy radio sources drawn from the three
largest deep 1.4 GHz radio surveys (0852$+$17 field, Condon \&
Mitchell 1984; 1300$+$30 field, Mitchell \& Condon 1985; 0846$+$45
field, Oort 1987). All the selected sources have radio fluxes greater
than 0.1 mJy and optical counterparts brighter than $V = 20.0$
(0846$+$45 field), $B = 21.4$ (0852$+$17 field) and $B = 22.3$
(1300$+$30 field). The radio flux limit varies across the survey and
the areal coverage ($\Omega(S)$) of the B93 sample is a complicated
function of radio flux, which is plotted in Figure \ref{fig:omega}.
Spectra have been obtained for 87 of these, providing object type and
spectroscopic redshifts, though at $z>0.35$ H$\alpha$ is unobservable
in their spectra so the spectroscopy is only complete at low
redshifts.  Of these sources with spectra, 47 turned out to be
star-burst galaxies. The B93 sample is the largest spectroscopic
sample of sub--mJy sources so far available in literature and, due to
its radio and optical limits, is the most complete sample of
radio--selected sub--mJy star--forming galaxies.  This sample is
therefore the best currently available and is the most suitable for
estimating the star--formation history as traced by the radio
luminosity density of star forming galaxies.

In all the following analysis we use only those objects within B93
which Rowan-Robinson {\em et al.} identified as star-burst galaxies.

\section{The luminosity function and luminosity density}
\subsection{Basic Method}

To calculate both the luminosity function and the luminosity 
density we require an estimate of the effective volume within
which each object of the sample could have been observed.  
Besides a $z\leq0.35$ spectroscopic redshift limit, the B93
 sample has two flux limits, one radio and one optical,
this effective volume is thus given by the following expression
\begin{eqnarray}
V_i=
\int_0^{\infty} P( {\cal O}|L_i,L_{B,i},z )\Omega_{\rm max}\frac{dV}{dz} dz
\end{eqnarray}
where  $P( {\cal O}|L,L_{B},z )$ is the probability of observing
an object of radio luminosity $L$ and optical luminosity $L_B$ at
redshift $z$ within our survey, given by
\begin{eqnarray}
P({\cal O}|L_i,L_{B,i},z)=\sum\frac{\Omega(S(L_i,z))}{\Omega_{\rm max}}
   P({\cal O}|S_B),
\end{eqnarray}
where the summation is over each of the three survey areas, in our
case $P({\cal O}|S_B)$ is either 1 or 0 depending on
whether the magnitude is below or above the limit in that survey area.
The radio fluxes are calculated assuming a radio spectral index,
$\alpha=d\log(S_\nu)/d\log(\nu)=0.8$; the optical fluxes are
calculated using the $K$-corrections for Scd galaxies given by
Metcalfe {\em et al.} (1991)
and we assume 
a cosmology of $\Omega_0=1$, $\Omega_\Lambda=0$, $H_0=50h_{50}$ km
s$^{-1}$ Mpc$^{-1}$ throughout. 

Having determined $V_i$ we can use the estimator $\varphi (L)=\sum
1/V_i$ (Schmidt 1968), and likewise estimate the luminosity density by
$\Phi_{1.4}= \sum L_i/V_i$.  In the absence of luminosity cuts these
estimators are unbiased, however one might worry about using these
estimators to determine the radio luminosity function and luminosity density
if the optical flux cut was much more severe than the radio, leading
to poor sampling of the radio luminosity plane.  We will
demonstrate that this is not the case.
 
In the presence of luminosity cuts, introduced for example by 
redshift cuts these estimators need to be modified thus:

\begin{eqnarray}
\Phi_{1.4}=
\frac{1}{C({L_{\rm min},L_{B,{\rm min}}})}
\sum \frac{L_i}{V_i}\label{lumden}
\end{eqnarray}

where $C({L_{\rm min},L_{B,{\rm min}}})$ is a measure of the ``completeness''
of the sample being the fraction of luminosity density above the
minimum observable radio and optical luminosities
i.e.
\begin{eqnarray}
C=\frac
{\int_{L_{\rm min}}^{+\infty}\int_{L_{B,{\rm min}}}^{+\infty}L \varphi(L,L_B) d \log L_B d\log L}
{\int_{-\infty}^{+\infty}\int_{-\infty}^{+\infty}L \varphi(L,L_B) d \log L_B d\log L}.\label{eqn:compl}
\end{eqnarray}
$\varphi(L,L_B)$ is the 1.4 GHz optical bi-variate luminosity function.

We will show that these correction factors are principally determined
by the shape of the luminosity function.  Maximum likelihood
estimators (e.g Sandage {\em et al} 1979) can determine the shape of
the luminosity function independently of density variations which can
strongly effect  the $1/V$ estimator.  Using a method akin to the maximum
likelihood technique we take some reasonable a priori models for
$\varphi(L)$ (and selection effects given by $P_i({\cal O}|L)$) to
predict the expected luminosity distribution
independently of density fluctuations, which can be compared with the
observed luminosity
distribution using a K-S test.
The expected luminosity distribution for an object in the
survey is given by
\begin{eqnarray}
f_i(L)=\frac{P_i({\cal O}|L)\varphi(L)
}{\int_{-\infty}^{+\infty}P_i({\cal O}|L) \varphi(L) d\log L}\label{ldist}.
\end{eqnarray}
The expected luminosity distribution of the sample is then given
by $\sum f_i(L)$.

\subsection{Impact of optical flux limit}

\begin{figure}
\epsfig{file=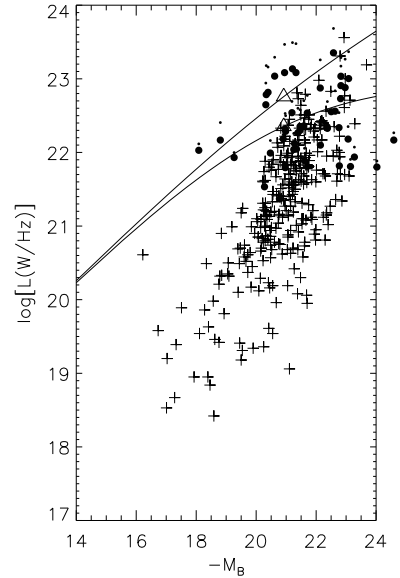, width=12cm}
\caption{Optical Luminosity versus Radio Luminosity for the Condon
sample (crosses) and for the B93 sample (filled circles) and
for the B93 sample corrected to $z=0$ luminosity according to
$L=L(z=0)(1+z)^{3.1}$ (filled squares).  Over-plotted is the locus of the
radio and optical flux limits for different redshifts, no evolution
upper, de-evolved to $z=0$, lower}\label{fig:optrad}
\end{figure}

 To demonstrate that the optical selection effects are less
significant than the radio selection effects in Figure
\ref{fig:optrad} we plot the radio and optical luminosities of the
Condon (1987, C87) spirals, which has $>99\%$ complete radio detections in
a sample of Revised Shapley-Ames galaxies (see Condon 1987 for more
details). Note that there is effectively only one flux limit, so this
correlation is not a ``distance vs. distance'' selection effect.

The $z\leq0.35$ star-bursts from B93 are also plotted in Figure
\ref{fig:optrad}, as is the passage of its double flux limits to
$z=0.35$.

The immediate point to notice from this Figure is that the
$L,L_B$ plane is not well sampled by the B93 sample,  the low
radio luminosity region is poorly sampled.  However this
is due to the small volume at low redshift, there appears to be
reasonable  sampling across $L_B$ for a give radio luminosity.

For a non-evolving population, virtually all the radio sources lie above 
our optical limit at $z\leq 0.35$. In other words, if we regard our 
sample as a purely radio flux limited sample, then we can be confident of 
very high completeness to the radio flux limit. Rowan-Robinson {\em et al.}
demonstrated that luminosity evolution of the same strength 
as required for IRAS galaxies $L(z)=L(z=0)(1+z)^{3.1}$ was
sufficient to explain the sub-mJy radio counts.
Such a  pure
luminosity evolution is equivalent to a large negative radio
$K$-correction term, and the corrected positions in Figure
\ref{fig:optrad} also demonstrate a high completeness.

We can quantify this by using the radio-optical correlation to
estimate the incompleteness as a function of redshift, then use the
observed redshift distribution to get a first-order estimate of the
overall incompleteness:
\begin{equation} 
f = 100\% \times N_{\rm obs} / \sum k(L_{{\rm min},i}, z_i) = 98.4\% 
\end{equation}
where $f$ is the completeness, $N_{\rm obs}$ the size of the
star-burst sample, $k$ is $1$ over the fraction of the radio-optical
correlation above the optical flux limit as a function of redshift and
luminosity, $L_{{\rm min},i}$ is the faintest radio luminosity
observable at redshift $z_i$, and the sum is performed over the
star-burst sample. The B93 sample does appear to have a slightly
larger dispersion in the radio-optical plane than the local sample,
but this larger scatter still yields $96.7\%$ completeness. We
conservatively assume no passive stellar evolution, the effect of
which would be to increase our completeness further still.  N.B. this
measure of ``completeness'' is not the same as that described in
Equation \ref{eqn:compl}.  The assertion that the optical selection
does not have a significant additional impact on the radio selection
is consistant with the findings from deep optical observations of the
Marano Field (Gruppioni {\em et al.} in preparation) that the faintest
optical identifications of sub-mJy radio sources do not appear to be
star-burst galaxies.

\subsection{$1/V$ Luminosity Function}

\begin{figure}
\epsfig{file=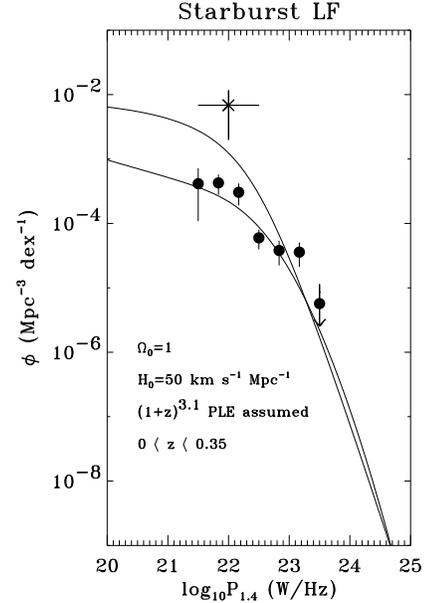,width=12cm}

\caption{The $1/V$ Luminosity function for the B93 sample: filled
circles. Also
illustrated is the luminosity function from Condon 1987,
upper curve,
the warm IRAS luminosity function of S90, lower curve.
Also illustrated is the estimate of luminosity function 
from Windhorst (1985)}\label{fig:radlf}
\end{figure}

The luminosity function is plotted in Figure \ref{fig:radlf}, which
agrees with the calculation in Rowan-Robinson et al. (1993).  It is
clear from Figure \ref{fig:radlf} that the faint end of the luminosity
function is not well-constrained by this sample.  This is due to the
small volume at low redshift of this sample.  For $z<0.05$ (where the
limiting radio luminosity is $L=1.1\times 10^{21} h_{50}^{-2}{\rm WHz}^{-1}$
the B93 sample covers only $2700 h_{50}^{-3}Mpc^{3}$, and using the
S90 ``warm'' luminosity function we would only expect 1.75 galaxies
within this volume.  This small volume at low redshifts is also
very susceptible to large scale structure variations.  Assuming
the power spectrum of Peacock and Dodds (1994) we would expect
RMS density fluctuations of around 60 to 60 per cent
over such a volume (allowing for the survey being spilt into three).

The co-moving number densities are consistent with the single star-burst
at $z<0.35$ detected in the Hubble Deep Field (Richards et al. 1998),
as well as with the luminosity function from the $z<0.35$ Marano
star-bursts (Gruppioni et al 1998, in preparation). This data 
is however in strong disagreement with the very high star-burst density
claimed in the micro-Jansky radio population by Windhorst et
al. (1995). Figure \ref{fig:radlf} plots 
the $z<0.35$ Windhorst et al. star-bursts, which have an order of
magnitude higher space density than our estimate. The nature
of the $\mu$Jy population is controversial however, with Hammer et
al. (1995) claiming the population is dominated by weak AGN, albeit in a
sky area which may be biased by the presence of a cluster. Part of the
discrepancy may also be due to high frequency ($5-8.5$ GHz)
selection in both the Hammer and Windhorst samples, where the radio
spectral index may not be steep due to the 
flat-spectrum contributions from H{\sc ii} regions and possibly AGN. 
The high-frequency selected sub-mJy/$\mu$Jy 
population may therefore be being sampled higher up
the luminosity function, or may be contaminated by Seyfert galaxies. 
This justifies our choice of a low-frequency selected
sub-mJy radio sample, albeit in retrospect. 

\subsection{Testing the shape of the luminosity function}

We have already discussed that the B93 sample is insufficient to
determine $\varphi(L)$ at faint radio luminosities
 since the volume and thus number of
objects at low redshift is small.

To estimate the corrections to our estimate of the luminosity density
(Equation \ref{lumden}) we are only interested in the shape of
the luminosity function
(where shape includes a measure of the typical Luminosity
e.g. $L_{*}$) and not the overall normalisation which cancels in
Equations \ref{eqn:compl} and \ref{ldist}.  We thus wish to
test the shape of B93 luminosity function against some
reasonable a priori functions.

The IRAS luminosity function is well determined even at the faint end,
it is natural to test an IRAS luminosity function modified via the
FIR/radio correlation (Helou, Soifer \& Rowan--Robinson 1985). For a
galaxy with the model star-burst spectrum of Efstathiou,
Rowan--Robinson \& Seibenmorgen (1998, in preparation) this relation
can be translated to
\begin{equation}
S_{60~\mu m}~ =~ 133~ S_{1.4~GHz}\\
\end{equation}
or 
\begin{equation}
L_{60~\mu m}~ =~ 133~ L_{1.4~GHz}\\
\end{equation}

We consider two models, the first ($l1$) is the ``warm'' luminosity
function of S90(their solution 25), (modified so $L_{*}=133 L_{60*}$).
It is appropriate to use the ``warm'' sample, since these are the
actively star-forming galaxies, similar to this sample.  This
luminosity function is also clearly similar in shape and normalisation
to our observed radio luminosity function (Figure \ref{fig:radlf}).
The use of this model luminosity function is independent of the
overall normalisation, and assumes the shape is constant with
redshift. It is known that IRAS galaxies strongly evolve (e.g. S90,
Oliver {\em et al 1994}, etc.), and if this evolution only changes the
normalisation of the luminosity function then this is
immaterial. However if shape of the luminosity function was to change
with redshift, this would affect our estimate of the luminosity
density.  Neither IRAS samples, nor this sample alone are sufficient
to determine what sort of evolution is appropriate to these types of
samples; however the faint radio counts do not support pure density
evolution as strong as that suggested in S90 (Rowan-Robinson {\em et
al.} 1993).  Thus it is appropriate to consider some model with an
evolution of the Luminosity with redshift.  Our second model $l2$
assumes the same zero redshift luminosity function as $l1$ but has
$L=L_{0}(1+z)^{3.1}$.  Since the S90 $z=0$ luminosity function was not
calculated with this evolution, this model is not entirely consistent
and will over-predict the numbers of luminous galaxies, and lead to an
under-correction to the Luminosity density, nevertheless the two
models together  give a reasonable idea of the level of
uncertainty in the luminosity function, and would bracket a more
natural model where Luminosity evolution was included in the estimate
of the Luminosity function self consistently.

Our third model, $l3$,
assumes the shape of the low $z$ Condon {\em et al} (1987) radio
luminosity function, with no evolution.

In addition to the luminosity functions we 
consider a number of different models for the effect of the joint optical
and radio limits on the observable radio luminosities $P_i({\cal O}|L)$:
\begin{eqnarray}
P_i({\cal O}|L)=\frac{\Omega(S(L,z_i))}{\Omega_{\rm max}} \label{m0}\\
P_i({\cal O}|L)=\frac{\Omega(S(L,z_i))}{\Omega_{\rm max}}
   \frac{\Omega_B(S_B(L L_{B,i}/L_i,z_i))}{\Omega_{B, \rm max}}\label{m1}\\
P_i({\cal O}|L)=\frac{\Omega(S(L,z_i))}{\Omega_{\rm max}}\label{m2}
    \frac{\int_{\infty}^{+\infty}\Omega_B P(L_B|L)dlgL_B}{\Omega_{B, \rm max}}
\end{eqnarray}

(In the above equations $B$ represents the optical which may be either
$B$ or $V$ band, where corresponding magnitudes of individual objects
are computed using the $S_{5000}/S_{7000}$ ratio in B93 and magnitude
limits are adjusted using the average $m_B-m_V=0.7$.)  Model $m0$
(Equation \ref{m0}) assumes that the optical magnitude of the object
is totally unrelated to the radio luminosity so the only selection
that comes into play is the radio flux limit.  If we knew the radio
flux limit as a function of position Equation \ref{m0} would reduce to
$P_i({\cal O}|L)=0, L<L_{{\rm min},i}; P_i({\cal O}|L)=1, L\ge L_{{\rm
min},i}$.  Equation \ref{m1} (model $m1$), assumes that the lower
luminosity objects potentially missed through the optical selection
have the same radio/optical colour as the object under consideration,
this will be satisfactory if the colours within the sample are
representative of the colours in the underlying population (in fact
the optical and radio flux limits will survey to restrict the range of
colours); in our case of simple magnitude limits $\Omega_B/\Omega_{B,
\rm max}=0, S_B<S_{B, \rm lim}; s \Omega_B/\Omega_{B, \rm max}=1, S_B
\ge S_{B, \rm lim}$.  Model $m2$ (and $m3$) take the form of Equation
\ref{m2} and require some estimate for the marginal optical radio
distribution function $P(L_B|L)$: model $m2$ takes this from Saunders
{\em et al.} (1990, S90), assuming $L_{60}=133 L$; model $m3$ takes
this from the Condon {\em et al} (1991).

Using the different models ($m0,m1,m2,m3$) for the optical and radio
selection criteria $P_i({\cal O}|L)$ and the three models for the
luminosity function $l1,l2,l3$ we calculate the luminosity distributions
and compare with the observed luminosity distributions using a
K-S test, these results are summarised in Table~\ref{kstest}.
(In this and subsequent analysis we have excluded the 5 objects
which fall below the optical completeness limits quoted in Benn {\em
et al.}.)
From this we can see that none of the models tested can be
significantly ruled out.   There is very little difference
between the models which assume either no optical selection 
effects or optical selection effects determined from a priori
estimates of the radio/optical distribution,  this is because
these estimates do not predict any significant loss of sources
to the optical selection effects.  A better fit is provide by
models which assume the optical selection is determined by the
distribution of colours within the sample itself ($m1$).  All forms
of luminosity function give acceptable fits, assuming $m1$.

\begin{table}
\begin{tabular}{lccccc}
$P_i({\cal O}|L)$ & $m0$ & $m1$ & $m2$ & $m3$  \\
$\varphi(L)$         & $l1$ & $l1$ & $l1$ & $l1$  \\
$P(KS)$           & 0.13 & 0.64 & 0.13 & 0.13   \\
\\
$P_i({\cal O}|L)$ & $m0$ & $m1$ & $m2$ & $m3$  \\
$\varphi(L)$         & $l2$ & $l2$ & $l2$ & $l2$  \\
$P(KS)$           & 0.30 & 0.83 & 0.29 & 0.30   \\
\\
$P_i({\cal O}|L)$ & $m0$ & $m1$ & $m2$ & $m3$  \\
$\varphi(L)$         & $l3$ & $l3$ & $l3$ & $l3$  \\
$P(KS)$           & 0.04 & 0.33 & 0.04 & 0.04   \\

\end{tabular}
\caption{Comparison of the observed and expected luminosity
distributions under various assumptions about the underlying 
luminosity function and selection effects.}\label{kstest}
\end{table}

\begin{figure}
\epsfig{file=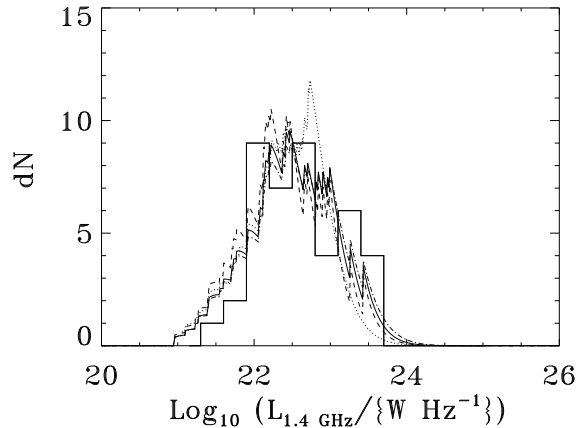,angle=90,width=8cm}
\caption{The observed and expected luminosity distributions. Solid
histogram, observed distribution. Dotted curve expected luminosity
distribution assuming ``warm'' IRAS luminosity function ($l1$) and
no accounting for optical selection ($m0$, similar to $m2,m3$).  Solid
curve $l1$ and optical selection determined by objects colours
($m1$). Dot-dashed curve, evolving ``warm'' IRAS luminosity function
($l2$).  Dashed curve Condon Luminosity function ($l3$).}

\end{figure}

\subsection{The luminosity density}

The correction factor 
($C({L_{\rm min},L_{B,{\rm min}}})$, Equation \ref{eqn:compl}) to be applied to the luminosity density
in the presence of luminosity (i.e. redshift) cuts,
can be re-expressed and 
estimated in a number of ways, we choose to express it as
\begin{eqnarray}
C&=&\frac
{\int_{L_{\rm min}}^{+\infty}L \varphi(L)  d\log L}
{\int_{-\infty}^{+\infty}L \varphi(L) d\log L}\nonumber\\
&-&\frac
{\int_{L_{\rm min}}^{+\infty}\int_{L_{B,{\rm min}}}^{+\infty}L \varphi(L,L_B) d \log L_B d\log L}
{\int_{-\infty}^{+\infty}\int_{-\infty}^{+\infty}L \varphi(L,L_B) d \log L_B d\log L}.
\end{eqnarray}

The last term in this equation represents the loss in luminosity
density due to the optical cuts alone.  We have calculated this term
at $z=0.05$ and $z=0.175$
using the Condon bivariate luminosity function, the Saunders bivariate
luminosity function, and for a version of the  Condon bivariate
luminosity function with increased dispersion appropriate for the
B93 sample.  The largest value we get for the second term is 
${\cal O} 10^{-5}$,  which is negligible compared to the former term involving
only the radio luminosity density ${\cal O} 1$.  Since the 
normalisation cancels out, the only significant correction to be
applied is due to the shape of the luminosity function.

With surveys of small volume density fluctuations on the scale of the
survey volume can contribute significantly to the uncertainty in
derived quantities.  To estimate the errors due to this cosmic
variance we determine the effective wavenumber $k$ appropriate for a
cube with volume one third of the total volume in the redshift range,
since the survey is composed of three independent volumes.  From this
we can estimate the density fluctuations $\sigma^2=\Delta^2(k)$ from
the power spectrum of Peacock and Dodds (1994), these we reduce by a
factor of three, assuming the three separate areas are independent.
These errors are approximate and do not take properly into account the
survey geometry or the fact that lower luminosity objects sample
smaller volumes, all these effects would serve to increase the errors.
These errors are slightly larger than the shot noise terms to which
they are added in quadrature.  We tabulate the cosmic variance
errors for this and other samples used for estimating luminosity densities(Table \ref{tab:lss}).

\begin{table}
\begin{tabular}{llccll}

Sample   & Area    &sub. & $z_{\rm min}$, $z_{\rm max}$ & $k_{\rm eff}/h_{50}$ & $\sigma$ \\
         & $/10^{-4}$sr     &                   &               & /${\rm Mpc}^{-1}$ &  \\
B93:     & 3.4  & 3& 0,      0.05 & 0.30   & 0.72 \\
B93:     & 3.4  & 3&0.05,    0.175& 0.095  & 0.41 \\
B93:     & 3.4  & 3& 0.175,  0.35 & 0.055  & 0.26 \\
B93:     & 3.4  & 3& 0.05,   0.35 & 0.052  & 0.25 \\
B93:     & 3.4  & 3&0,       0.35 & 0.052  & 0.25 \\
Tresse \&& 0.42  & 5& 0,     0.3  &  0.040 & 0.16 \\
 Maddox: \\
Lilly    & 0.42 & 5&   0.2,   0.5 &   0.028& 0.10 \\
Lilly    & 0.42 & 5&   0.5,   0.75&   0.024& 0.088\\
Lilly    & 0.42 & 5&   0.75,   1  &   0.023& 0.079\\
Gallego  & 1400 & 1&   0,   0.045 & 0.0089 & 0.036\\
Gronwall & 210  & 1& 0,      0.085& 0.0092 & 0.017 \\

\end{tabular}

\caption{Cosmic variance errors ($\sigma=\sqrt{\Delta^2(k)}$) within
surveys of different areas, redshift ranges and different number of
subdivisions. For the B93, Tresse \& Maddox and the low $z$ Lilly
samples these errors are comparable with the shot noise errors.}
\label{tab:lss}
\end{table}

In Table \ref{tab:lumden} we tabulate the un-corrected
luminosity density for the B93 sample over 4 redshift ranges.
We also tabulate the correction factors that need to be applied
assuming various luminosity functions which we have shown to
be compatible with this sample,  (for the evolving luminosity
function we have take then luminosity function shape to be 
defined at roughly the centre of the redshift bin).
We also tabulate the luminosity density estimated from a
Schecter function fit to the luminosity function
Condon RSA excluding AGN (Serjeant et al, in preparation).
We also tabulate the luminosity density for warm IRAS galaxies
from S90, scaled by $\Phi_{60\mu{\rm m}}=133\Phi_{1.4}$.

\begin{table}
\begin{tabular}{ccrrr}
 $z$& 
$\log \Phi_{1.4}$ &
\multicolumn{3}{c}{$-\log C(L_{\rm min},B_{\rm min})$}\\
&$/ h_{50}{\rm WHz}^{-1}{\rm Mpc}^{-3}$&$l1$&$l2$&$l3$\\
  0-0.05     &   $19.27  \pm  0.14 $ & 0.0  & 0.0  & 0.0\\
  0-0.35     &   $19.09  \pm  0.14 $ & 0.0  & 0.0  & 0.0\\
  0.05-0.35  &   $19.09  \pm  0.14 $ & 0.03 & 0.02 & 0.05\\
  0.05-0.175 &   $18.91  \pm  0.20 $ & 0.03 & 0.02 & 0.05 \\
  0.175-0.35 &   $19.01  \pm  0.16 $ & 0.21 & 0.12 & 0.42\\
\\
  0-0.1      &   $18.69  \pm  0.05 $ & 0.0  & 0.0  & 0.0\\
  0-0.1      &   $19.08  \pm  0.04 $ & 0.0  & 0.0  & 0.0\\

\end{tabular}
\caption{Estimates of the raw 1.4GHz luminosity density 
in different redshifts ranges and corrections that need to
be applied under different assumptions
about the underlying luminosity function.  The $0<z<0.05$
luminosity density is taken from Serjeant {\em et al.} (in preparation).  The last two rows are
obtained from converting the
``warm'' and total IRAS luminosity densities of S90 to a 1.4Ghz luminosity density
using $133 L_{1.4}=L_{60}$.
}\label{tab:lumden}
 
\end{table}

We can see from this that the total luminosity density in the B93
sample as a whole is in good agreement with the luminosity density
from the Condon sample.  The luminosity density in the higher redshift
bin is larger than that in the lower redshift bin by a factor of
between 1.7 and 3.7, depending on the luminosity function applied,
consistent with luminosity evolution at a rate of $(1+z)^{3.1}$,
however, due to the cosmic variance this is not very
significant.

The luminosity density from the low-redshift sample alone is a factor
of around two lower than the Condon sample and similar to the
luminosity density as estimated from the ``warm'' IRAS galaxies.  Due
to the cosmic variance the discrepancy with the Condon
sample is not significant, while the discrepancy between the Condon
sample and the ``warm'' IRAS sample is.  It is likely that the Condon
sample also includes the radio equivalent of the IRAS ``cool'' Cirrus
galaxies. 

Using the $H\alpha$ luminosities quoted in Benn {\em et al} 1993,
corrected for slit loss using the APM magnitudes and continuum fluxes
we can use the same effective volumes to estimate the $H\alpha$
luminosity density within this sample, Table \ref{tab:halpha}.  This may be an underestimate
of the total $H\alpha$ luminosity density since from some objects
$H\alpha$  is not detected and we have not included the 
spectroscopic limits in our effective volume.

\begin{table}
\begin{tabular}{cl}

$z$  & $\Phi(H\alpha)$ \\
     & $/ h_{50}{\rm W}Mpc^{-3}$ \\
 0.0-0.35   & 31.8627 \\
 0.05-0.35  & 31.7764 \\
0.05-0.175  & 31.7554\\
0.175-0.35  & 31.2052\\

\end{tabular}
\caption{$H\alpha$ Luminosity density estimated from 
star-forming galaxies within the B93 sample.  Errors and correction
factors will be similar to those in \protect\ref{tab:lumden}}\label{tab:halpha}
\end{table}

\section{Conversion from Luminosity Density to star formation rate}

Nearly all of the radio emission from star--forming galaxies is 
synchrotron
radiation from relativistic electrons and free-free emission from H{\sc ii}
regions. Only massive stars (i.e. $M~ \gsimeq~ 5 M_{\odot}$) ionise the 
H{\sc ii} regions and produce supernovae, whose remnants accelerate most of the
relativistic electrons. Such massive stars have lifetimes much shorter 
than
the Hubble time, so the current radio luminosity is proportional to the
recent star formation rate (Condon 1992). The supernova rate is directly
related to the non--thermal radio luminosity, which at $1.4$ GHz is 
dominant
over the thermal component (see Condon \& Yin 1990 for the most reliable
derivation of this relation, calibrated with Galactic supernova 
remnants). 
Moreover, since all stars more massive than $M = 5 M_{\odot}$ become radio
supernovae, the radio supernova rate is determined directly by the star
formation rate.  Thus, the star formation rate can be obtained by
non--thermal radio luminosity, following Condon (1992): 
\begin{equation}
SFR(M \geq 5 M_{\odot}) = \frac{L_{1.4~GHz}~[W Hz^{-1}]}{5.3 \times
10^{21} (\frac{\nu}{GHz})^{\alpha}} ~M_{\odot} yr^{-1} 
\end{equation}
where $\nu$ is the frequency and $\alpha$ is the non--thermal radio 
spectral
index (as defined above). If we assume a Salpeter initial mass function 
(IMF; $\psi(M)\propto M^{-2.35}$, $0.1-125$ $M_\odot$) we obtain for
the total star formation rate
\begin{equation}
SFR(M \geq 0.1 M_{\odot}) = \frac{L_{1.4~GHz}~[W Hz^{-1}]}{7.63\times10^{20}}
\end{equation}

\begin{figure}
\epsfig{file=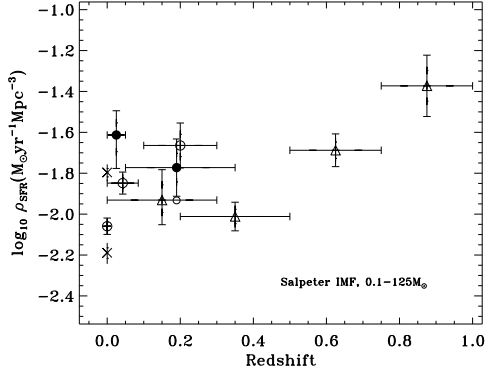,width=10cm}
\caption{The co-moving star-formation rate $z<1$ as estimated from:
Filled circles = 1.4GHz (Left to right: Condon, B93)
Open circles = $H\alpha$ (Gallego, Gromwell, B93, Tresse \& Maddox)
Open triangles = UV (Treyer, Lilly, Lilly, Lilly).
Corrections for reddening have not been made or have been removed}
 \label{fig:sfr1}
\end{figure}

\begin{figure}
\epsfig{file=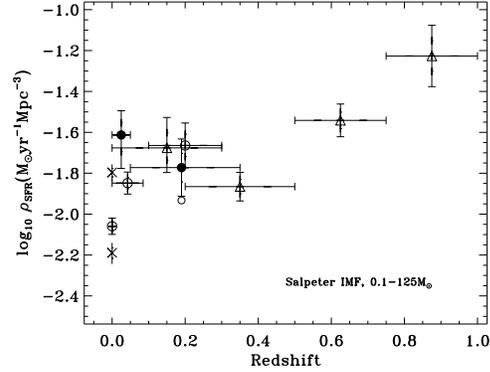,width=10cm}
\caption{The co-moving star-formation rate $z<1$ as estimated from:
Filled circles = 1.4GHz (Left to right: Condon, B93)
Open circles = $H\alpha$ (Gallego, Gromwell, B93, Tresse \& Maddox)
Open triangles = UV (Treyer, Lilly, Lilly, Lilly)
Corrections for reddening have been made or retained} \label{fig:sfr2}
\end{figure}

In Figures \ref{fig:sfr1}, \ref{fig:sfr2} we convert our radio
luminosity densities (Table \ref{tab:lumden}, row 3, correction $l1$) to star 
formation rates together with that from the RSA sample (Table \ref{tab:lumden}, row 1).

It is instructive to compare with the U-band luminosity 
densities at redshifts $z\stackrel{<}{_\sim}1$, from Treyer et al. (1998) 
and
the Canada-France Redshift Survey (CFRS, Lilly et al. 1996).
For the 
ultraviolet conversions we assume 
\begin{equation}
SFR(M \geq 0.1 M_{\odot}) = \frac{L_{2000\AA}~[W]}{7.94\times10^{20}}
\end{equation}
\begin{equation}
SFR(M \geq 0.1 M_{\odot}) = \frac{L_{2500\AA}~[W]}{6.7\times10^{20}}
\end{equation}
taken from Treyer et al. (1998) and Cowie et al. (1997)
respectively. 
In Figure \ref{fig:sfr1} we do not apply a reddening 
correction, in Figure \ref{fig:sfr2} the luminosity densities
are de-reddened following Treyer et al. (1998).

Also plotted are the H$\alpha$ luminosity densities from the CFRS
(Tresse \& Maddox 1997) the KISS sample (Gromwell 1998) 
and from the local ($z=0$) sample of Gallego
et al. (1995), converted to total star formation rates for this IMF
using
\begin{equation}
SFR(M \geq 0.1 M_{\odot}) = \frac{L_{H\alpha}~[W]}{1.41\times10^{34}}
\end{equation}
from Madau et al. (1996). Both H$\alpha$ luminosity densities were
corrected for reddening using Balmer decrements and assuming a simple
dust screen, in Figure \ref{fig:sfr1} we remove the average reddening
correction $\sim1$ mag (Tresse \& Maddox 1997).

A priori we would have expected the 1.4GHz to provide a higher
estimate of the star-formation rate than an un-reddened optical
estimate.  This is because the radio should be measuring the
total star-formation rate which the optical only measures the
un-obscured rate.  At the low redshift ($0<z<0.05$) this 
does indeed appear to be the case, the Condon luminosity density
is well above that estimated from the $H\alpha$ of Gallego {\em et al.} (1995)
and Gromwell (1998), and above the UV estimate of Treyer {\em et al.}
(1998).  At intermediate redshifts ($0.05<z<0.35$), probed by
the B93 survey,  the Tress \& Maddox $H\alpha$ estimate from the
CFRS and the UV estimate of Treyer {\em et al.} (1998), the estimates
agree,  if the reddening corrections are applied then the optical
estimators can even exceed the radio estimate (Figure \ref{fig:sfr2}).  
This is initially surprising until we notice that the errors are
large,  we note that we have applied the cosmic variance terms to
all the data points and these are significant, particularly for
the B93 sample and the Tresse \& Maddox sample.  We thus suspect
that the coincidence of these various estimators may be due to
differences in the mean density of these survey volumes which
masks a difference in their sampling of the star-formation rate.
This assertion is strengthened when we see that the $H\alpha$ estimate
of the star-formation rate from the B93 sample itself is further below
the radio estimate than the Tresse \& Maddox point is above it.

An additional star-formation indicator is the far infrared.  Several
authors (Scoville \& Young 1983; Thronson \& Telesco 1986; Condon
1992; Rowan--Robinson et al. 1997) have derived relations to convert
from 60 $\mu$m luminosity to star formation rate.  These calculations
assume some proportion of the optical luminosity produced by young
stars is absorbed and re-emitted in the infrared.  (This same
extinction factor should naturally be applied to the star-formation
rates calculated from the UV.)  From such calculations it is of course
possible to deduce the FIR/Radio correlation. Condon (1992) deduced
the FIR/Radio correlation assuming an obscuration factor of 2/3.,
recently Cram et al.  (1998) seems to indicate a close agreement
between the radio SFR estimate and the far--infrared one obtained
considering $\sim100$ per cent reprocessing of starlight in the
far-infrared.  In general it does appear that the extinction fraction
is high.  It would in fact be possible to use the FIR/Radio
correlation to determine the extinction fraction for any assumed
underlying IMF, this could then be feed self consistently into the UV
and $H\alpha$ SFR estimates.  Such an analysis is beyond the scope of
this paper.

Instead we use the FIR/Radio correlation simply to convert local
estimates of the $60\mu$m luminosity density from S90 to 1.4GHz
luminosity density and then use the 1.4GHz calibration above which
is independent of extinction.  We plot in Figure
\ref{fig:sfr1},\ref{fig:sfr2} the star-formation rate estimated thus
from the S90 luminosity densities for both 
``warm'' IRAS galaxies and for all IRAS galaxies.
It is not clear that the ``cool'' IRAS galaxies trace star-formation
as do the warm IRAS galaxies, since the cool emission is from 
cirrus clouds which may be illuminated by older stellar populations
(similarly the UV density of quiescent or weakly star-forming galaxies
might also be dominated by old stars).  

\section{Conclusion}

We have estimated the 1.4 GHz luminosity density from a sample of
radio star-burst galaxies $z<0.35$, demonstrating that the optical
selection criteria applied to this sample are not important to this
determination.  The small volume of this survey at $z<0.05$ prevents
us from having accurate estimates of the luminosity density from low
radio luminosity star-forming galaxies.  Nevertheless we demonstrate
that a number of reasonable a priori luminosity functions can be used
to estimate the luminosity density missed.  These corrections are
small for the sample as a whole, though larger and more disparate for
a higher redshift sub-sample.  The 1.4 GHz luminosity density
increases as we move to higher redshifts, though the strength of this
effect is dependent on the assumed luminosity function and may not be
conclusive with a sample of this volume.
Overall the 1.4 GHz luminosity density agrees with that obtained from
1.4 GHz measurements of the RSA sample of Condon {\em  et al.}.  

The 1.4GHz estimator of star-formation rates should be unaffected by 
extinction,  whereas local estimates from the UV and $H\alpha$ should
be.  We were thus surprised to find the estimate from the radio 
broadly consistent with these other estimates, particularly since
the local estimate from the Condon luminosity density was considerably
higher than from optical measures.  Since the luminosity density
estimated from $H\alpha$ within the B93 sample produces a much 
lower estimate of the star-formation rate than the 1.4GHz estimate
and the (optimistic) cosmic variance errors in the B93 and CFRS samples are
comparable to the shot noise errors we attribute this apparent
agreement to clustering fluctuations.   We thus strongly suggest that 
spectroscopic follow-up of larger area radio surveys is needed to
accurately determine the un-obscured star-formation rates $0<z<0.3$.

The star formation rates at these redshifts do not conflict with the
Far-IR background measurements, since the sub-mJy radio population
make a negligible contribution to the FIR background. At these
redshifts the $1.4$ GHz and $850\mu$m fluxes are roughly equal, and
the Puget et al. (1996) $850\mu$m background corresponds to about $26$
Jy per square degree.

\section*{Acknowledgements}

\label{lastpage}

\end{document}